\documentclass[10pt,conference]{IEEEtran} 

\usepackage{cite}
\usepackage{graphicx}
\usepackage{amsmath,amssymb}
\usepackage{algorithm}
\usepackage{algorithmic}
\usepackage{booktabs}
\usepackage{url}
\usepackage{xcolor}

\newcommand{\orcid}[1]{\href{https://orcid.org/#1}{\includegraphics[scale=0.06]{orcid.pdf}}}

\usepackage[colorlinks,urlcolor=blue,linkcolor=blue,citecolor=blue]{hyperref}
\expandafter\def\expandafter\UrlBreaks\expandafter{\UrlBreaks\do\/\do\*\do\-\do\~\do\'\do\"\do\-}
\usepackage{array}
\usepackage{multirow}

\usepackage{booktabs}

\usepackage{amssymb}
\usepackage{xcolor}
\newcommand{\greencheck}{{\color{green!45!black}\checkmark}}

\begin{document}


\title{TinyAC: Bringing Autonomic Computing Principles to Resource-Constrained Systems}

\author{
  \IEEEauthorblockN{Wojciech Kalka\IEEEauthorrefmark{1}, 
                   Ruitao Xue\IEEEauthorrefmark{2},
                   Kamil Faber\IEEEauthorrefmark{1},
                   Aleksander Slominski\IEEEauthorrefmark{3},\\
                   Devki Jha\IEEEauthorrefmark{2},
                   Rajiv Ranjan\IEEEauthorrefmark{2},
                   Tomasz Szydlo\IEEEauthorrefmark{2}}
  \IEEEauthorblockA{\IEEEauthorrefmark{1}AGH-UST, Krakow, PL}
  \IEEEauthorblockA{\IEEEauthorrefmark{2}Newcastle University, Newcastle upon Tyne, UK}
  \IEEEauthorblockA{\IEEEauthorrefmark{3}IBM TJ Watson, NY, USA}
}








\maketitle

\begin{abstract}
Autonomic Computing (AC) is a promising approach for developing intelligent and adaptive self-management systems at the deep network edge. In this paper, we present the problems and challenges related to the use of AC for IoT devices. Our proposed hybrid approach bridges bottom-up intelligence (TinyML~+~on-device learning) and top-down guidance (LLMs) to achieve a scalable and explainable approach for developing intelligent and adaptive self-management tiny systems. Moreover, we argue that TinyAC systems require self-adaptive features to handle problems that may occur during their operation. Finally, we identify gaps, discuss existing challenges and future research directions.
\end{abstract}


The motivation behind applying the principles of \textit{Autonomic Computing (AC)} to low-power Internet of Things (IoT) devices stems from the growing complexity, unreliability, and scalability issues of current IoT deployments. Many IoT systems fail in the real world due to factors such as poor integration, unreliable connectivity, inadequate power management, limited hardware resources, inconsistent data quality, and a lack of security and real-time monitoring. In addition, interoperability challenges across heterogeneous devices, insufficient lifecycle planning, and the absence of adaptive mechanisms lead to unsustainable systems that cannot evolve or scale efficiently.

\begin{figure}
\centerline{\includegraphics[width=16.5pc]{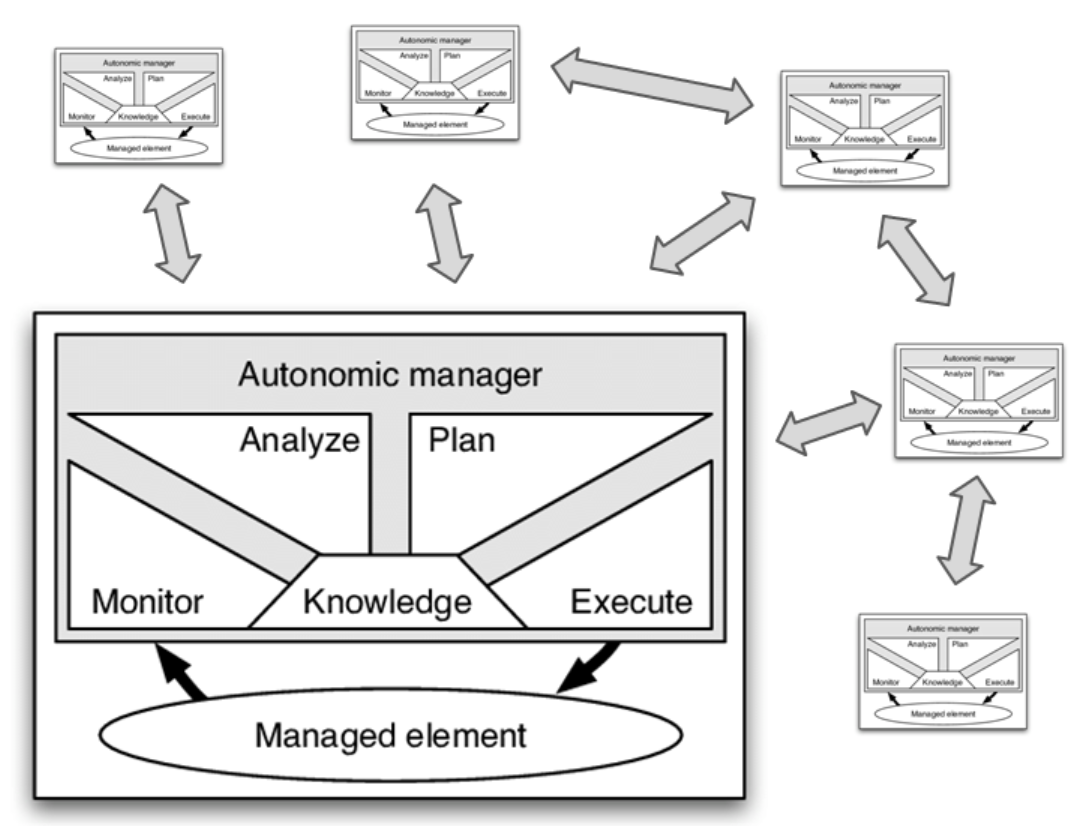}}
\caption{Autonomic Computing agents}\vspace*{-5pt}
\label{fig:ac-vanilla}
\end{figure}

For many years, researchers worked on \textit{Autonomic Computing}, a self-management paradigm inspired by the human autonomic nervous system, aimed at reducing the complexity of managing increasingly distributed, dynamic, and heterogeneous computing systems. Introduced by IBM (Fig.~\ref{fig:ac-vanilla}) in the early 2000s, the paradigm defines systems that can automatically monitor themselves, analyse internal and external conditions, plan responses, and, finally, execute appropriate adaptations with minimal human intervention. 


Although this vision is still difficult to achieve in its full form, recent advances in LLM, explainable AI, and neurosymbolic machine learning make the realisation of this vision more tangible \cite{AC_LLM}. However, pushing these principles toward billions of resource-constrained smart Internet of Things devices 
introduces new challenges, such as limited compute, memory, and energy, requiring lightweight and efficient implementations of autonomic behaviours.

The heart of modern, typically battery-powered, IoT devices is the microcontroller, which integrates the essential components of a computing system into a single, compact chip designed for real-time, low-power, and cost-sensitive embedded applications. These devices usually have less than 2MB of Flash memory and less than 1MB of RAM. Microcontrollers units (MCUs) are extremely cost- and power-sensitive yet now powerful enough to support TinyML, enabling on-device intelligence. 

TinyML, short for Tiny Machine Learning (ML), is a subset of ML that is optimised to run on very low-power MCUs targeting Edge or Embedded use cases.
The size of the TinyML market is expected to reach \$10B in 2033\footnote{\url{ https://www.verifiedmarketreports.com/product/tiny-machine-learning-tinyml-market/}}.

The global production of microcontroller units is substantial, with annual shipments reaching approximately 24.6 billion units this year. This figure reflects a significant increase from previous years, driven by the growing demand for MCUs in various applications, including automotive systems, industrial automation, consumer electronics, and the Internet of Things (IoT)\footnote{\url{https://www.marketresearchfuture.com/reports/microcontroller-unit-market-8626}}.

Despite the intelligence provided by TinyML, these devices remain largely static. The operational context of IoT systems, such as battery level, CPU load, network conditions, and environment, demands context-aware adaptation to maintain efficiency, safety, and robustness. Without such adaptability, IoT devices risk delivering incorrect results when environmental conditions deviate from their original design assumptions. However, there are techniques and protocols to remotely manage the fleet of IoT devices \cite{Micro_Szydlo}, but these devices still lack autonomy. Our purpose is to propose the \textit{Tiny Autonomic Computing} framework for IoT devices, discuss its required functionalities, and identify open challenges.





\section{Understanding Moore's Law}
The progress of CPU development can be described using Moore’s Law, which is the observation that the number of transistors on a microchip doubles approximately every two years, leading to increases in computing power and decreases in relative cost. However, its influence on the microcontroller market differs from that on general-purpose processors, due to the specific constraints and goals of MCU (e.g., low power, real-time performance, and cost-effectiveness). Unlike desktop CPUs, microcontroller designs do not always chase maximum transistor counts. Instead, they must strike a balance between power consumption, cost, and long-term availability.

\begin{table}[h]
    \caption{Key Feature Additions Across Different Eras}
    \centering
    \begin{tabular}{>{\raggedright\arraybackslash}p{0.25\linewidth}>{\raggedright\arraybackslash}p{0.6\linewidth}}
        \hline
        Era & Key Feature Additions \\
        \hline
        1970s–1990s & Basic ALU, single-core, pipelining, early ISAs \\
        1990s–2000s & Caching, FPUs, branch prediction, SIMD \\
        2000s–2010s & SoC integration, GPUs, DSPs, power management \\
        2015–present & NPUs, AI acceleration, secure enclaves, vector instructions \\
        Emerging & TinyML optimizations, on-device training, IMC, RISC-V extensions \\
        \hline
    \end{tabular}
    \label{tab:key_feature_additions}
\end{table}

Shrinking the size of the transistor reduces the dynamic and leakage power, enabling ultra-low-power MCUs. It has significant design implications for battery-powered and energy-harvesting applications, such as IoT, wearables, and wireless sensors. 
More powerful cores (e.g., Cortex-M7, RISC-V-based MCUs) can be designed without significantly increasing power consumption. MCUs can now run more complex algorithms, enabling TinyML, digital signal processing, and real-time analytics directly on-device. Table~\ref{tab:key_feature_additions} presents the key feature additions to the MCUs introduced over the years.

With more transistors available on the same die space, microcontrollers can integrate more features (e.g., ADCs, DACs, timers, and communication interfaces like SPI, I2C, CAN, and USB), making these devices more versatile but also more complex to manage. Also worth mentioning is the inclusion of wireless communication interfaces in the modern SoC. \textit{Low Power Wide Area Network} (LPWAN) and \textit{Personal Area Network} (PAN) technologies provide IoT connectivity but differ in range, topology, and use cases. LPWAN options like LoRa, Sigfox, NB-IoT, and LTE-M support long-range, low-power communication, while PAN technologies such as Zigbee, Thread, and BLE enable short-range, low-latency mesh communication ideal for smart homes and industrial networks.

Finally, more silicon space allows hardware-based security features (e.g., TrustZone, secure boot, encryption engines), which are essential for secure IoT deployments, especially in industrial and medical domains.

\section{Smart IoT devices --- TinyML}
TinyML is rapidly gaining popularity because it aligns with key trends in AI, IoT, edge computing, sustainability, and real-time intelligence. It enables intelligent sensing, autonomous decision-making, and ubiquitous AI where traditional ML is infeasible due to constraints in power, compute, or connectivity. 

We can observe the dynamic development of algorithms for TinyML, ranging from image analysis using DNN, through 3D motion pattern recognition, to audio analysis. Virtually every major microcontroller manufacturer offers appropriate SDKs (Software Development Kits) supporting TinyML~\cite{tinyml_survey} or has acquired a company specialising in this area, including the recent Edge Impulse acquisition by Qualcomm or Neuton.AI acquisition by Nordic Semiconductor.

To compare the solutions and their optimisations, the benchmarks such as \textit{MLPerf Inference Tiny benchmark suite} \cite{mlperf} measure how fast systems can process input and produce results using a trained model. Despite the increasing possibilities of running ML models on devices with very limited resources, which makes them increasingly smart, the devices still have limited self-adaptability.

\section{Context is Key}
What distinguishes IoT systems from other IT solutions is the operational context in which they operate. The operational context and the application domain 
are essential in IoT as they directly influence how a device should behave, communicate, and consume resources. Context provides critical situational awareness, ensuring efficient, safe, and intelligent operation. IoT devices often have limited power, compute, and bandwidth. Operation context (e.g., battery level, CPU load, network conditions) will enable the device to make informed trade-offs.


Smart devices driven by ML models are designed to operate in a specific environment and, as a consequence, may produce unexpected results when used under different conditions. For example, a video-based occupancy sensor developed for smart home applications might not work correctly in the high-temperature and humid environment of the factory. 
Therefore, IoT devices require a high level of autonomy that allows them to adapt and operate efficiently in evolving conditions.


\section{Autonomic Computing}
Autonomic Computing is often referred to as the MAPE loop, consisting of \textit{Monitor}, \textit{Analyse}, \textit{Plan}, and \textit{Execute} components, all supported by a shared \textit{Knowledge} base. Fig.~\ref{fig:ac-vanilla} depicts the classical architecture as proposed in the original IBM paper.

Autonomic systems are characterised by four key self-* properties:
\begin{itemize}
    \item \textit{Self-Configuration} – the ability to configure and reconfigure automatically under changing environments.
    \item \textit{Self-Optimization} – the ability to continuously improve performance and resource efficiency.
    \item \textit{Self-Healing} – the ability to detect, diagnose, and recover from faults.
    \item \textit{Self-Protection} – the ability to anticipate and defend against malicious attacks or failures.
\end{itemize}

In a more general sense, autonomous agents have the ability to interact with each other and collaborate toward a common goal. The solutions in this case are much more complex as they often require the exchange of knowledge between agents.

\section{AC \& Resource Constraints}
Designing an Autonomic Computing Manager (ACM) for IoT devices based on MCUs requires integrating the MAPE-K loop (Monitor, Analyze, Plan, Execute, Knowledge) with resource-aware, energy-efficient, and real-time capabilities of embedded operating systems.

\begin{figure}
\centerline{\includegraphics[width=18.5pc]{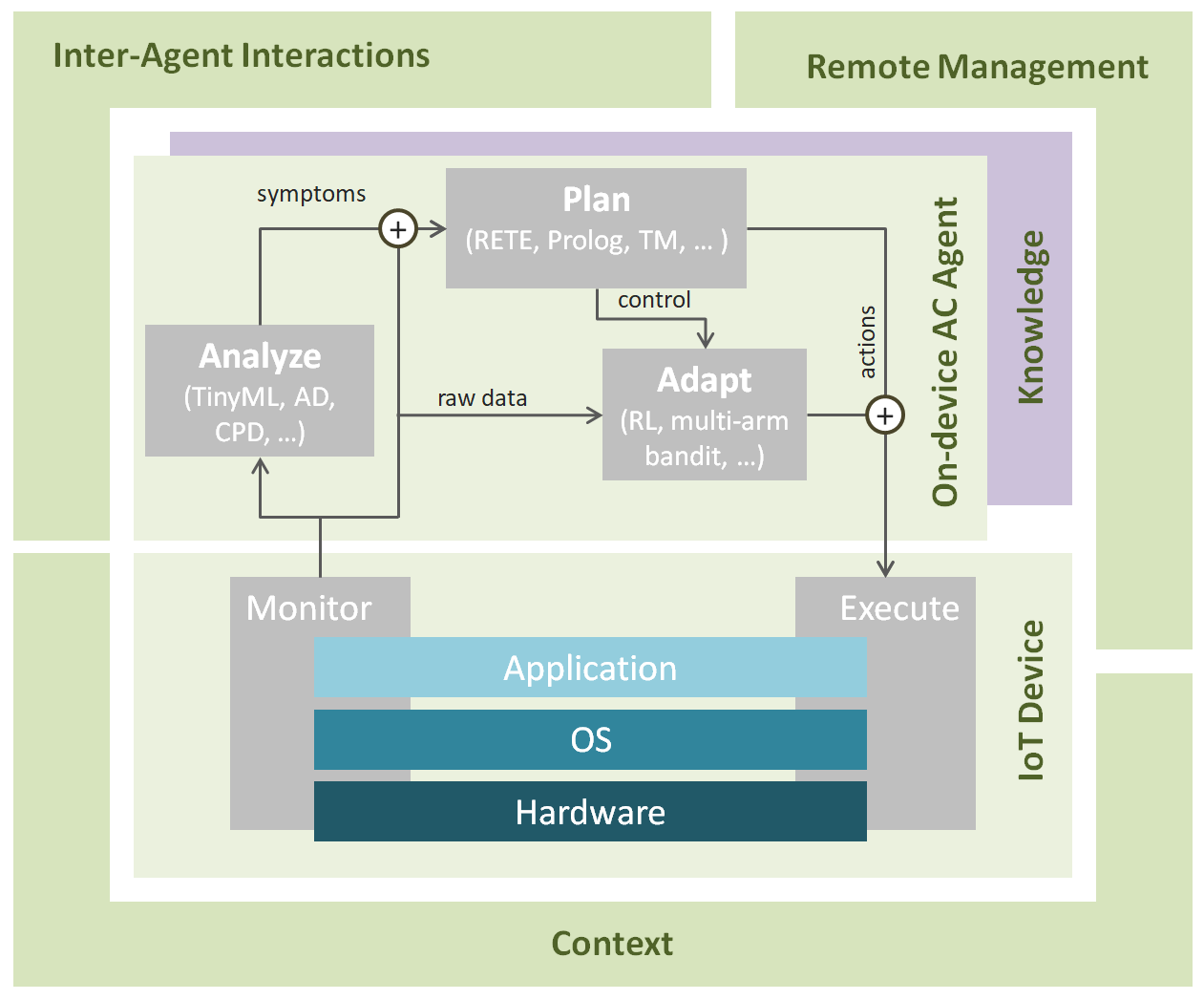}}
\caption{Concept of Tiny Autonomic Computing Agent for IoT devices}\vspace*{-5pt}
\label{fig:tiny-agent}
\end{figure}

Due to the strict resource constraints of MCU-based IoT devices, the algorithms used in the MAPE loop should be employed selectively, depending on the system goals, resource availability, and the complexity of the application. Figure~\ref{fig:tiny-agent} depicts the concept of Tiny AC Agent. In the \textit{monitor} stage, the raw data should be collected from the external context through embedded sensors, hardware components, as well as the operating system and the application itself. Examples of such data include time series values corresponding to CPU utilisation, temperature, and accelerometer data. In the \textit{analyze} stage, raw data should be converted to symptoms reflecting high-level situations, such as spikes in CPU utilization or identification of device failure. The \textit{plan} stage is extended by \textit{adapt} to provide adaptation mechanisms driven by methods such as rule-based policy systems or reinforcement learning. Finally, the \textit{execute} stage is an enforcement point for the actions issued by the plan and adapt stages. The \textit{knowledge} in this case supports all the algorithms used in the autonomic loop, storing their internal policies, historical data, system models, and context.

Table~\ref{tab:rtos} shows the standard and emerging algorithms for each MAPE component, with a focus on those suitable for TinyML and edge AI systems.

\begin{table*}[h]
    \caption{Example algorithm selection for MAPE-K Components}
    \centering
    \begin{tabular}{>{\raggedright\arraybackslash}p{0.2\linewidth}p{0.8\linewidth}}
        \toprule
        \textbf{Tiny MAPE-K Components} & \textbf{Algorithms and information sources} \\
        \midrule
        \textbf{Monitor} & Sensor APIs ($I²C$/SPI), sensor\_samples, real-time system telemetry hooks, thread metrics, stack usage \\
        \textbf{Analyze} & TinyML model, anomaly detection, change point detection, anomaly detection \\
        \textbf{Adapt} & Reinforcement learning algorithms, multi-arm bandit  \\
        \textbf{Plan} & Decision policies (state machine, rule engine) \\
        \textbf{Execute} & Power modes, scheduler hints, runtime configuration e.g. threads priority \\
        \bottomrule
    \end{tabular}
    \label{tab:rtos}
\end{table*}

\begin{figure}
\centerline{\includegraphics[width=18.5pc]{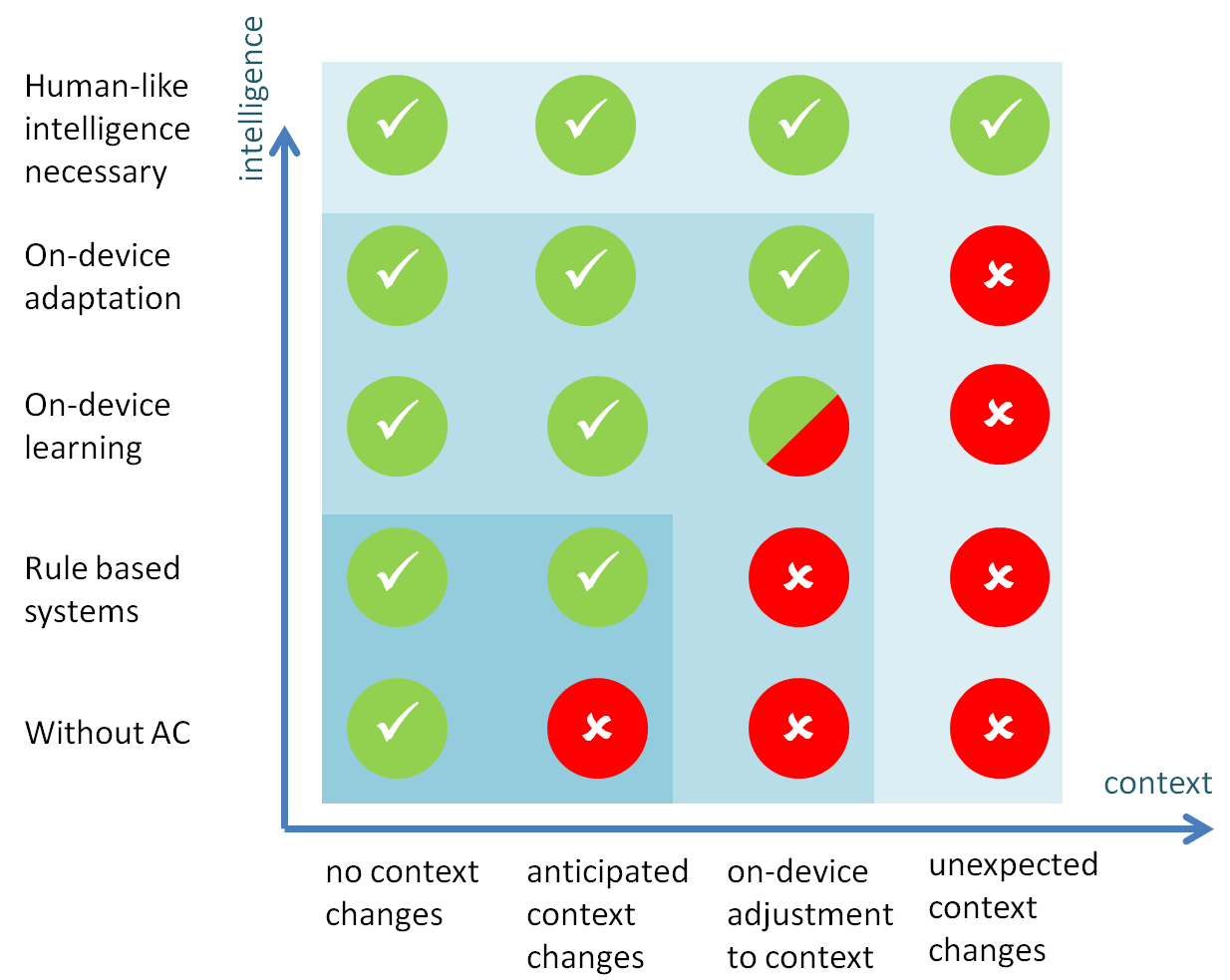}}
\caption{Adaptability to context changes}\vspace*{-5pt}
\label{fig:ac-intelligence}
\end{figure}

\section{Why On-Device Learning and Adaptation for AC?}
Providing \textit{on-device learning} and \textit{on-device adaptation} is necessary to enable true AC in IoT systems, particularly for applications requiring real-time, context-aware, and personalised adaptation. 
\textit{On-device learning} contributes to each stage of MAPE loop in unique and indispensable ways. 
Figure~\ref{fig:ac-intelligence} illustrates the relationship between system intelligence levels and their ability to handle varying context changes. It shows five categories of intelligence --- from \textit{Without AC} to \textit{Human-like intelligence} --- and four levels of context variability from \textit{no context changes} to \textit{unexpected context changes}. Green check marks indicate the ability to handle context changes, while red crosses indicate failure when faced with such conditions. 

For applications where simple adaptability to context changes is expected, the classic \textit{rule-based} systems are sufficient. For changes that are more complex but still within the same application domain, the \textit{on-device adaptation} is necessary to infer symptoms based on the analysis of the sensor data. For example, if the background noise of the signal changes, the rapid detection of this change may trigger the follow-up actions leading to the agent adapting to a new environment. Table~\ref{tab:detectors_comparison} provides the comparison of the $pre-$ and $post-$ deployment detectors that might be used in such situations. 

More complex cases require adjustments to the context situation, which involves observing the environment and the effect of the performed action. This leads us to the \textit{on-device adaptation} to the context, which can be implemented leveraging techniques such as reinforcement learning. An example would be a smart camera equipped with an array of LED spotlights to illuminate a scene. In a setup where some light sources are obscured, the AC agent can interact with the environment and learn which light sources should be turned off to preserve energy.

Still, in all of these situations, the changes in context are within the envisaged ranges. Only systems with \textit{human-like} intelligence can reliably manage unexpected context changes that go beyond what was envisioned at the algorithm design stage. 
 
%

\begin{table*}[h]
    \caption{Comparison of \textit{predeployment} and \textit{postdeployment} symptom detectors}
    \centering
    \begin{tabular}{>{\raggedright\arraybackslash}p{0.2\linewidth}>{\raggedright\arraybackslash}p{0.3\linewidth}>{\raggedright\arraybackslash}p{0.3\linewidth}}
        \toprule
        \textbf{Feature}         & \textbf{Predeployment Detectors}                                          & \textbf{Postdeployment Detectors}                                      \\ \midrule
        \textbf{Timing}          & Before device is deployed                                                 & After device is deployed                                               \\ 
        \textbf{Goal}            & Prevent defective or misconfigured devices from entering service          & Maintain performance, detect issues in real-world conditions           \\ 
        \textbf{Environment}     & Controlled (lab, factory)                                                 & Uncontrolled (real-world/field), adjusted after device deployment                                       \\ 
        \textbf{Focus}           & Hardware integrity, calibration, config validation                        & Fault detection, performance degradation, adaptive recovery            \\ 
        \textbf{Tools Used}      & Factory test rigs, diagnostic software               & On-device monitors, remote diagnostics, anomaly detection systems, change-point detection and data drift algorithms    \\ \bottomrule
    \end{tabular}
    \label{tab:detectors_comparison}
\end{table*}

\section{LLM Role in AC for IoT}
Large Language Models (LLMs) can play several enabling and complementary roles in Autonomic Computing for TinyML and IoT systems, even though LLMs themselves are typically too large to run on-device. Their role is best envisioned as part of the edge-cloud continuum, where LLMs can serve as knowledge engines, orchestrators, or planners that enhance or coordinate the autonomy of distributed, resource-constrained systems. Table~\ref{tab:llm_role} shows the role of LLM at various maturity levels taken from the original IBM's work on AC and applied to AC for tiny IoT devices.

At \textit{ level 0}, the devices are not supported by AC at all. The deployment of TinyML models does not guarantee that IoT devices will have autonomous computing features. \textit{Level 1} means that the device has basic functionalities and, to some extent, might be monitored. The role of LLM in this case is limited to suggesting AC algorithm selections and generating configuration code. There is no tight interaction with the devices. \textit{Level 2} assumes remote monitoring of devices leveraging protocols such as LwM2M~\cite{Micro_Szydlo}. Based on the error logs, LLM can generate the problem description in natural language, helping device management engineers in their work. \textit{Level 3} requires the ability to predict potential device problems, analyse various situations, and generate fallback scenarios for device recovery. At \textit{Level 4}, the LLM generates adaptation rules for the devices, taking into account situational awareness and potential problems. \textit{Level 5} assumes full autonomic device management, including device reprogramming, remote troubleshooting, and problem solving.

The higher the maturity level of LLM for IoT systems, the more their knowledge should be appropriately tailored to encompass the expertise of technicians and programmers responsible for creating such solutions. The use of Small Language Models (SLM) in this area is interesting, as this would enable the implementation of autonomous IoT islands, where individual devices could exchange knowledge with each other without constant cloud connectivity, which might be essential for managing devices in hard-to-reach places such as LEO micro satellites.


\begin{table*}
    \caption{AC Autonomy levels and the role of Large Language Models}
    \centering
    \begin{tabular}{>{\centering\arraybackslash}p{0.05\linewidth}>{\centering\arraybackslash}p{0.1\linewidth}>{\centering\arraybackslash}p{0.4\linewidth}>{\centering\arraybackslash}p{0.4\linewidth}}\toprule
         Level&  Description&  Description& LLM Role\\\midrule
         0&  Manual&  No autonomic capabilities; fully human-managed& N/A\\
         1&  Basic&  Basic monitoring and alerting& Design assistant; TinyML pipeline suggestions\\
         2&  Managed&  Sensors collect metrics; human still decides& Natural-language log analysis; report summarization\\
         3&  Predictive&  System predicts and alerts on future events& Suggest adaptation triggers; early planner support\\
         4&  Adaptive&  System acts based on predictions or changes& Generates runtime policies; contextual reconfiguration\\
         5&  Autonomic&  Fully self-managing, self-learning, and evolving& Evolves knowledge base; autonomous planner; natural-language control\\ \bottomrule
    \end{tabular}
    \label{tab:llm_role}
\end{table*}


\section{USE CASE}
\begin{figure}
\centerline{\includegraphics[width=18.5pc]{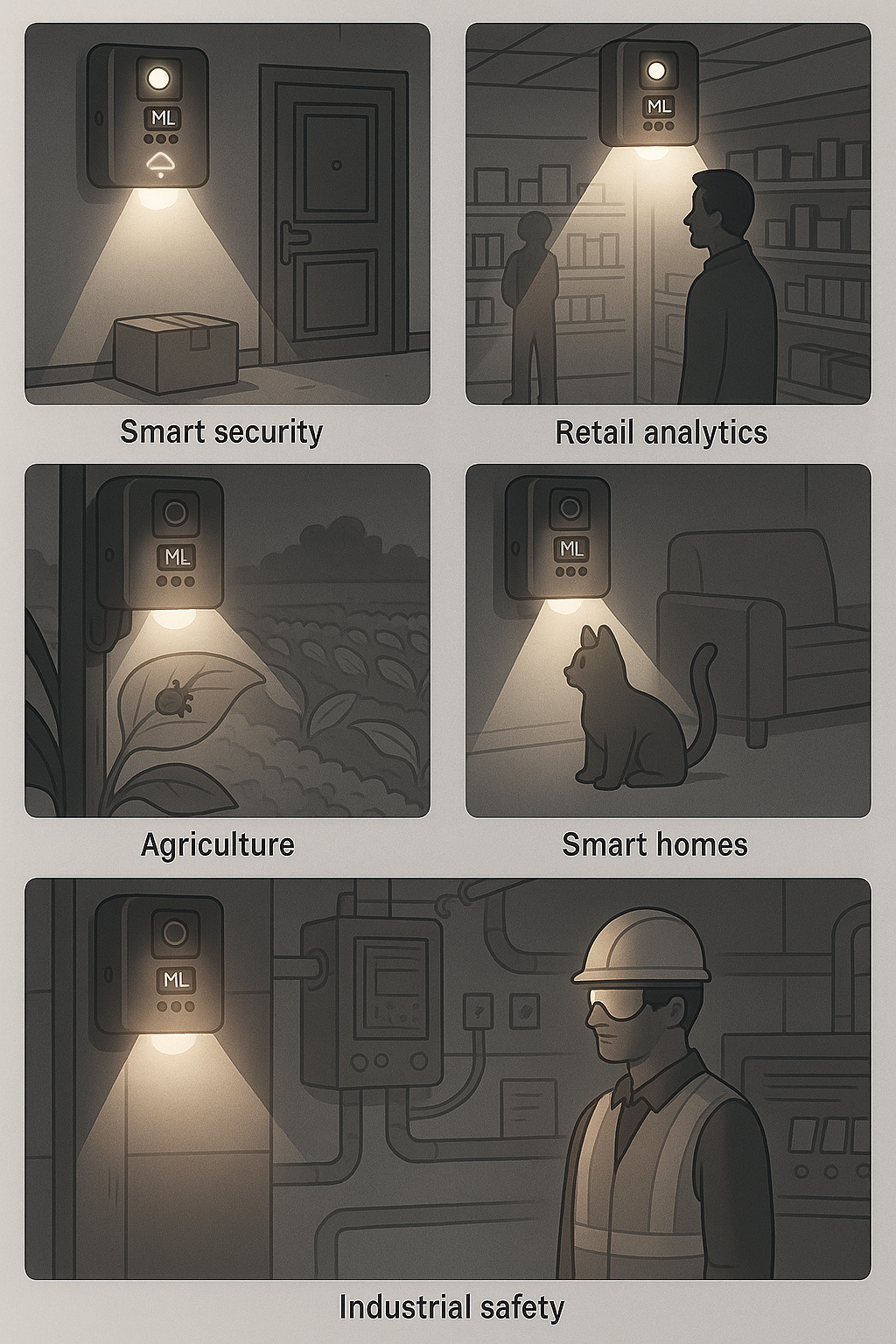}}
\caption{Operational context of IoT device (generated using  ChatGPT)}\vspace*{-5pt}
\label{fig:device}
\end{figure}

Let us consider an IoT device, which is a compact, low-power video analysis unit equipped with a tiny camera and an embedded machine learning model. The device captures images in real time, performs on-device inference to classify visual data (e.g., detecting objects, identifying motion, or recognising specific events), and transmits only the inference results to the cloud, minimising bandwidth and preserving privacy (Figure~\ref{fig:device}). Designed for edge intelligence, the device supports continuous operation in resource-constrained environments and might be useful in various application domains, including:
\begin{itemize}
    \item Smart security (e.g., intrusion detection, package monitoring)
    \item Retail analytics (e.g., customer counting, behaviour analysis)
    \item Agriculture (e.g., pest detection, crop monitoring)
    \item Industrial safety (e.g., hazard detection, worker presence tracking)
    \item Smart homes (e.g., pet monitoring, appliance usage tracking)
\end{itemize}

Although the main functionality of a smart camera is image processing using the dedicated machine learning model, the problems of ensuring its correct operation depend on its application and deployment. Table~\ref{tab:use-cases} illustrates the various situations that a device must face during its operation and the level of adaptability required to handle those situations. In the following subsections, we present selected examples that demonstrate the pragmatics of the tiny AC agent's operation.

\begin{table*}
\caption{Symptoms derived from context analysis and appropriate adaptation actions for smart IoT camera device. In the \textit{self-} column \textit{c.}, \textit{h.}, \textit{p.}, o. means configuration, healing, protection and optimisation. Accordingly in the adaptability levels column, \textit{s.r.}, \textit{o.l.} and \textit{o.a.} mean \textit{simple rules}, \textit{on-device learning} and \textit{on-device adaptation}. }
\centering
\begin{tabular}{>{\raggedright\arraybackslash}p{0.12\linewidth}>{\raggedright\arraybackslash}p{0.1\linewidth}>{\raggedright\arraybackslash}p{0.35\linewidth}>{\raggedright\arraybackslash}p{0.02\linewidth}>{\raggedright\arraybackslash}p{0.02\linewidth}>{\raggedright\arraybackslash}p{0.02\linewidth}>{\raggedright\arraybackslash}p{0.02\linewidth}>{\raggedright\arraybackslash}p{0.03\linewidth}>{\raggedright\arraybackslash}p{0.03\linewidth}l}\toprule
\multirow{2}{*}{Context} & \multirow{2}{*}{Symptoms} & \multirow{2}{*}{Action} & \multicolumn{4}{c}{Self-} & \multicolumn{3}{c}{Adaptability levels} \\
\cmidrule(lr){4-7}\cmidrule(lr){8-10}
 &  &  & c. & h. & p. & o. & s. r. & o. l. & o. a. \\
\midrule
\multirow{3}{*}{\shortstack[l]{Lighting and \\ Visual Conditions}} & Low light / night-time & Switch to infrared or night vision mode; adjust exposure and sensitivity & \greencheck&  &  &  & \greencheck&  &  \\\cmidrule(lr){2-10}
 & Bright sunlight / glare & Adjust contrast, white balance, or use HDR processing &  &  &  & \greencheck&  &  & \greencheck\\\cmidrule(lr){2-10}
 & Weather effects (rain, fog, snow) & Enhance image processing to maintain object detection accuracy & \greencheck &  &  &  & \greencheck &  &  \\
\midrule
\multirow{3}{*}{\shortstack[l]{Environmental\\ Conditions}} & Temperature extremes & Adjust internal cooling/heating, or power down non-critical functions to prevent damage &  &  & \greencheck &  & \greencheck &  &  \\\cmidrule(lr){2-10}
 & Dust, humidity, or moisture & Enter protective or low-power mode if sensors detect environmental risk &  &  & \greencheck &  &  & \greencheck &  \\\cmidrule(lr){2-10}
 & Vibration or physical disturbances & Temporarily pause AI inference or recalibrate motion detection to avoid false alerts &  & \greencheck &  &  &  & \greencheck &  \\
\midrule
\multirow{2}{*}{\shortstack[l]{Connectivity\\ Conditions}} & Low bandwidth & Lower video resolution, transmit only metadata or compressed snapshots &  &  &  & \greencheck &  &  & \greencheck \\\cmidrule(lr){2-10}
 & Intermittent network & Buffer video locally, reduce streaming quality, or prioritise event-based uploads & \greencheck &  &  &  &  & \greencheck &  \\
\midrule
\multirow{3}{*}{\shortstack[l]{Security and\\ Threat Conditions}} & Suspicious behaviour detected & Increase frame rate, activate high-priority logging, or alert security personnel &  &  & \greencheck &  &  & \greencheck &  \\\cmidrule(lr){2-10}
 & Attempted tampering or blockage & Trigger alarms, switch to alternative angles, or lock down system &  &  & \greencheck &  & \greencheck &  &  \\\cmidrule(lr){2-10}
 & Cybersecurity alerts & Limit network access, switch to local processing, or encrypt transmissions &  &  & \greencheck &  &  & \greencheck &  \\ \bottomrule
\end{tabular}
\label{tab:use-cases}
\end{table*}

\subsection{\textit{self-configuration}}
The tiny AC agent on the smart IoT camera activates its on-board video quality detectors and environmental sensors to monitor environmental context such as temperature and humidity. Based on real-time analysis, the device dynamically reconfigures its video filters, including sharpening and color balance, to adapt to the specific conditions of the video scene. Humidity and temperature are used to calculate the dew point and, in the risk of condensation of water on camera lenses, activate internal heating. The adaptability enhances the performance of the embedded video recognition algorithms, improving accuracy and robustness over time without requiring manual reconfiguration.

\subsection{\textit{self-healing}}
After deployment, the tiny AC agent on the IoT device continuously monitors its environment using built-in sensors.  A dedicated data fault detector runs in parallel to the main functionality of the device, analysing the sensor input streams for anomalies such as spikes, dropouts, or gaps that may compromise model accuracy. When such faults are detected, the autonomic agent intervenes by isolating the faulty sensor by disabling it at the operating system level. This action prevents unreliable data from propagating through the end-to-end system. In response, the application switches to a predefined fallback behaviour, ensuring continuity and resilience without requiring manual repair or restart.

\subsection{\textit{self-protection}}
After deployment, the tiny AC agent on the IoT device continuously logs its operational context using post-deployment context detectors. A sudden and unusual change in this context, such as a location change, specific movement pattern detection, or usage, is identified as a potential theft scenario. In response, the tiny AC agent activates self-protection protocols, disabling non-essential functionalities to conserve battery and isolating system resources, leaving only connectivity services active. This ensures that the device remains accessible for remote monitoring, diagnostics, or recovery actions, thereby enhancing resilience and physical security without user intervention.

\subsection{\textit{self-optimisation}}
The tiny AC agent embedded in the IoT device continuously monitors multi-modal contextual inputs, such as RF signals, motion patterns, and ambient audio, to assess operational needs. Based on this context, it dynamically activates, deactivates, or scales TinyML models to match the current resource availability (e.g., CPU, memory, battery) and task urgency, parallelly observing the effects of taken actions and adjusting them accordingly. This adaptive model management ensures optimal performance, energy efficiency, and responsiveness, allowing the device to intelligently prioritise tasks and maintain quality of service under varying conditions without external intervention.

\subsection{\textit{what if all self-* fail?}}
If previous context adaptation methods fail and the device is unable to adapt to context changes with the necessary flexibility, external support will be necessary. In this case, the device should provide raw data, symptoms, and access to execution modules to an external AI agent via, for example, the \textit{Model Context Protocol} (MCP). The LLM-based agent can remotely perform a range of diagnostic actions, such as scanning network SSIDs, running a battery test, or deactivating some sensors. Then, after analysis, a set of new adaptation policies can be generated, and the device's knowledge can be modified accordingly.

\section{Future Research Directions}

The proposed hybrid system bridges bottom-up intelligence (TinyML + on-device learning and adaptation) and top-down guidance (LLMs) to realise scalable, explainable, and robust Autonomic Computing at the deep edge. However, several challenges must be addressed to implement  TinyAC's vision.

\textit{Increased complexity.} TinyAC adds new autonomous capabilities to IoT devices, and that leads to an increase in complexity, even potential for emergent behaviours that may make managing such devices harder. There is a need for observability and other tools that are specific to autonomic capabilities, which go beyond what the existing TinyML tooling currently supports.

\textit{Lightweight time-series compression algorithms for remote diagnostics.} Lightweight time-series compression algorithms face key research challenges, including balancing compression ratio with low computational and memory overhead, especially for resource-constrained IoT devices. Preserving critical temporal patterns during lossy compression without degrading downstream analytics or machine learning performance is difficult.

\textit{Autonomic Computing benchmarks for MCU-based IoT systems.}
Developing autonomic computing benchmarks for MCU-based IoT systems presents challenges such as defining standard, lightweight metrics that capture self-management behaviours (self-configuration, self-healing, self-optimisation, and self-protection) within tight resource constraints. Benchmarks must account for diverse hardware, real-time performance, energy efficiency, and robustness under dynamic conditions. Capturing system adaptation, fault tolerance, and decision latency in a reproducible, scalable way is difficult. Integrating these benchmarks with TinyML and sensor variability further complicates design and evaluation.

\textit{Lightweight on-device anomaly detection algorithms.}
Lightweight on-device anomaly detection algorithms face challenges in achieving high accuracy with minimal computational, memory, and energy overhead, which is essential for MCU-based IoT devices. Designing models that generalise well across noisy, non-stationary, and streaming time-series data is difficult under constrained training and storage conditions. Balancing detection sensitivity with false positive control, ensuring real-time performance, and enabling online or incremental learning are key challenges. Additionally, adapting to heterogeneous sensor modalities and maintaining robustness without frequent cloud interaction complicates deployment and evaluation.

\textit{Lightweight logic-based machine learning models.}
The planning module in the proposed TinyAC system should leverage logic-based approaches, such as rule-based systems, \textit{Prolog}, and \textit{Tsetlin Machines}~\cite{TM}, due to their high explainability and ease of human interpretation. However, existing implementations for low-power devices typically lack support for updating rules and logical constructs at runtime. This limitation yields for solutions that not only execute rules efficiently but also enable on-device rule learning and adaptation, thereby allowing the system to evolve in response to changing conditions without external intervention.

\textit{Lightweight on-device adaptation algorithms.}
There is a need for research on on-device adaptation algorithms that include ultra-lightweight RL techniques operating within the severe CPU, memory, and energy constraints of microcontrollers. Approaches should include quantised or binary RL models, sparse or event-driven policy updates, and memory-efficient exploration strategies. These methods should aim to enable real-time, continuous adaptation to changing environments without relying on cloud resources. Key challenges involve balancing model accuracy, stability, and responsiveness with minimal execution latency and energy consumption. Emerging directions integrate adaptive learning rates, experience compression, and hybrid local–remote strategies, ensuring that embedded devices can maintain performance while operating autonomously in dynamic, resource-limited IoT scenarios.

\textit{Multi-Agent Autonomic Computing algorithms.}
The communication capabilities of modern SoCs enable AC agents on small IoT devices to interact, allowing them to coordinate their actions toward common goals. Furthermore, research into decentralised federated learning (DFL) algorithms is essential, as it empowers AC agents to collaboratively exchange and refine knowledge without relying on a centralised server, thereby enhancing scalability, robustness, and privacy.

\textit{Knowledge exchange formats.}
Saving and exchanging knowledge acquired by ML algorithms learning on devices is crucial for their seamless interoperability. For classic ML models, such formats exist, such as ONNX. However, for lifelong anomaly detection or rule-based systems, such solutions do not exist. The proposed formats should be size-efficient and be designed to support the IoT communication protocols and interfaces adopted by the industry.

\textit{Small LLMs trained to be IoT technicians.} Current large language models are designed for general-purpose use and possess only partial domain knowledge for managing IoT devices. To address this gap, it is crucial to train language models enriched with the expertise of technicians who maintain large-scale IoT networks. Equally important is the development of small language models (SLMs) capable of running on edge gateways, enabling local, low-latency management of IoT devices without relying on constant cloud connectivity. LLM post-training techniques such as \textit{Supervised Fine-Tuning (SFT)} and \textit{Direct Preference Optimisation (DPO)} 
could be used for that purpose.

\section{CONCLUSIONS}
In this paper, we presented the problems and challenges related to the use of Autonomic Computing for IoT devices. Until now, AC has been the domain of large computer systems, where the availability of resources for adaptive algorithms has been taken for granted. In IoT systems, unfortunately, such an assumption cannot be made, and the algorithms must be adapted to real-world constraints. The presented hybrid approach bridges bottom-up intelligence (TinyML + on-device learning) and top-down guidance (LLMs) to ensure scalable, explainable, and robust Autonomic Computing at the edge.

Going further, in the IoT-Edge-Cloud continuum, cooperating multi-autonomic computing agents are essential for managing complexity, ensuring scalability, and enabling autonomic behaviours. Each agent autonomously controls its local environment, such as an IoT device, an edge node, or a cloud service, while collaborating with other agents to optimise performance throughout the system. This coordination supports distributed decision-making, task orchestration, and fault tolerance. Agents should collaborate to strike a balance among latency, energy, bandwidth, and computing resources. For instance, an edge agent can offload processing to the cloud or reroute tasks during network congestion. This cooperation ensures Quality of Service (QoS) and supports resilience in context-aware dynamic environments.


\def\refname{REFERENCES}



\vspace*{-8pt}




\end{document}